\documentstyle[amssymb,prb,aps,epsfig,multicol]{revtex}  
\begin{document}
\title{Comment on ``Structural Stability and Electronic Structure
for Li$_3$AlH$_6$"}
\author{D.J. Singh}
\address{Center for Computational Materials Science,
Naval Research Laboratory, Washington, DC 20375}
\date{\today}
\maketitle

\begin{abstract}
Density functional calculations of the electronic structure are
used to elucidate the bonding of Li$_3$AlH$_6$. It is found that this
material is best described as ionic, and in particular that 
the [AlH$_6$]$^{3-}$ units are not reasonably viewed as substantially
covalent.
\end{abstract}


\begin{multicols}{2}

Vajeeston and co-workers (VRKF, Ref. \onlinecite{ref1}),
recently presented a detailed
pioneering study of the electronic structure and structural
phase changes under pressure for Li$_3$AlH$_6$.
The complex hydrides, $A_x (M$H$_4)_y$, with $A$=Li, Na, K, Mg, Ca, Sr
or a mixture
of these and $M$=B or Al seem promising for H storage since they
contain very high weight percent H, and much of the H content can
be evolved at moderate temperatures.
\cite{bog1,gross,fichtner,zuttel}
Further, in 1997 Bogdanovic
and Schwickardi reported that with certain metal
additions, particularly Ti, NaAlH$_4$ can be cycled. \cite{bog1}
However, this has yet to be achieved with other members of
this family. Better understanding of the bonding, chemistry
and thermodynamics
of these hydrides is clearly needed in searching for other
cyclable members of this family. Understanding the properties of Li$_3$AlH$_6$
is of importance because of its place as the intermediate product
in the decomposition of LiAlH$_4$. In fact, NaAlH$_4$ follows a similar
decomposition pathway, with Na$_3$AlH$_6$ as an intermediate product.

Based on their calculations, VRKF conclude that ``energetic degeneration
of Al-$p$ and H-$s$ states together with the spatially favorable
constellation of Al and H favors covalent bonding between Al and H"
and that ``[AlH$_6$]$^{3-}$ forms distinct covalently bonded units".
Here results of density functional calculations,
showing this not to be the case are presented. It is argued
instead that Li$_3$AlH$_6$
is best viewed as an ionic solid, nominally Li$_3^+$Al$^{3+}$H$_6^-$.

The present calculations were done within
the local density approximation (LDA)
to density functional theory,
using the general potential linearized augmented planewave (LAPW) method
with local orbital extensions. \cite{singh-book,singh-lo}
The convergence was checked by doing calculations with various
basis set sizes and Brillouin zone samplings. The LAPW method uses
as decomposition of space into atom centered
spheres and an interstitial. The sphere radii are constrained by
the requirement that they be non-overlapping and that core states
do not significantly spill into the interstitial. H LAPW sphere radii ranging
from 1.1 $a_0$ to 1.55 $a_0$ were tested and the results were found to be
insensitive to this choice.

Combined synchrotron X-ray and neutron diffraction experiments have
shown that Li$_3$AlD$_6$ occurs in spacegroup $R\bar{3}$,
$a$=8.07117 \AA, and $c$=9.5130 \AA.\cite{exp-struct}
This structure is near cubic. The rhombohedral setting
has $a$=5.636 \AA, and rhombohedral angle 91.45$^\circ$.
Experimental and calculated structural coordinates in this setting are
given in Table \ref{table1}. As may be seen, these are very close,
and in particular the calculated Al-H bond length
in good agreement with experiment [1.736 \AA~ and 1.734 \AA~ (the same to
within the precision of the calculation)
{\em vs.} experimental values of 1.754 \AA~ and 1.734 \AA,
for H1-Al1 and H2-Al2, respectively].
In this setting the Al are at the corner and body centered positions and
are octahedrally coordinated by H; the Li occur near the
(1/2,0,1/4) positions, and are coordinated by distorted H octahedra.
This structure in itself is not particularly suggestive of covalency.

\begin{figure}[tbp]
\centerline{\epsfig{file=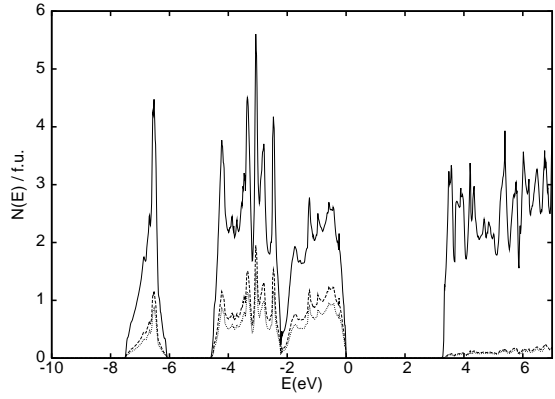,width=0.60\linewidth,angle=270,clip=}}
\vspace{0.3cm}
\caption{Electronic density of states [solid line] and $s$
projection onto the H
LAPW spheres, radius 1.3 $a_0$ (1.1 $a_0$),
[dashed (dotted) line] for Li$_3$AlH$_6$, in eV$^{-1}$
on a per formula unit basis using the experimental crystal structure.
The valence band maximum is at 0 eV.
Note the ionic nature shown by
the very different hydrogen $s$
contributions to the valence and conduction bands.}
\label{dos-exp}
\end{figure}

The calculated electronic density of states (DOS) and projections
onto the H LAPW spheres for the experimental crystal structure are
shown in Fig. \ref{dos-exp}. The DOS for the LDA atomic positions
is practically the same and agrees in its large features with
that of VRKF. It shows three distinct manifolds of
states. The two occupied manifolds are
a group of ``split-off" bands, containing a total of 2 electrons per
formula unit (4 per cell) at about -7 eV, relative to the valence
band maximum and a broader manifold containing 10 electrons per formula
unit. This latter manifold is almost split by a pseudogap (at -2 eV)
into a lower group of bands containing 6 electrons and an upper group
containing 4 electrons per formula unit. These valence bands are in
turn separated by a $\sim 3$ eV band gap from the conduction states.
The simplest covalent picture of an [AlH$_6$]$^{3-}$ unit would have
6 H 1$s$ orbitals, 1 Al $s$ orbital and 3 Al $p$ orbitals.
The Al $s$ orbital, if weakly hybridized, could form a ``split-off"
state holding 2 electrons per formula unit. The remaining orbitals,
would then form three manifolds of 6 electrons each: a lower bonding
group, a non-bonding group, and an upper antibonding group. Since
Li$_3$AlH$_6$ has 12 electrons per formula unit, the Fermi energy in
this scenario would intersect the non-bonding bands, yielding a metal
or perhaps a Mott insulating ground state. This is hardly consistent
with the calculated DOS. Furthermore, if the covalency were strong,
one would expect H character to be distributed between the three
$s$-$p$ manifolds. However, as may be seen from the projections of the
DOS, there is very much more H $s$ character in the valence bands
than in the conduction bands. Furthermore, the ``split-off" manifold
has approximately the same proportion of H $s$ character as the
main valence band manifold.

The simplest ionic picture would have full H$^-$ ions stabilized
by the Coulomb field.
In this picture, one would expect to have occupied
H $s$ derived valence bands accomodating the 12 electrons per formula
unit, separated by an insulating gap from metal derived conduction
bands.
Interactions between the 6 hydrogens in the octahedron,
which could be direct or via assisted hopping using unnoccupied
Al $sd$ orbitals, would break
the occupied H bands into a symmetric ($s$) combination (2 electrons),
a $t_{2g}$ combination (6 electrons) and an $e_g$ combination (4 electrons),
consistent with the shape of the DOS.
In this case, the valence DOS would be dominated by H $s$ character and
the conduction bands would have much less H $s$ character.

\begin{figure}[tbp]
\centerline{\epsfig{file=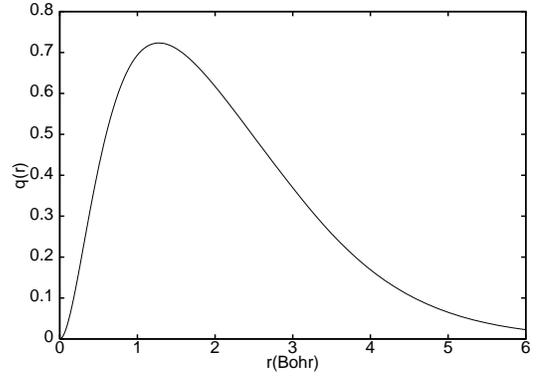,width=0.60\linewidth,angle=270,clip=}}
\vspace{0.3cm}
\caption{LDA charge distribution of a H$^{-}$ ion stabilized by a Watson
sphere (see text). The charge is defined as $q(r)=4 \pi r^2 \rho(r)$, where
$\rho(r)$ is the density and $r$ is the radius in Bohr.}
\label{hminus}
\end{figure}

To better assess the correspondance between this ionic scenario and
the calculated electronic structure,
we did atomic calculations in the LDA for H$^-$
stabilized by a Watson sphere, charge $+e$ and radius 3.06 $a_0$,
which is slightly smaller than the average H-Al distance (3.30 $a_0$).
The density profile of this ion is shown in Fig. \ref{hminus}.
A sphere of radius 1.1 $a_0$
would contain only 0.45 $e$ for this large ion, while
0.60 $e$ is in a 1.3 $a_0$ sphere.
As mentioned, it is not possible to use a sufficiently small Al LAPW
sphere to exclude H character and reflect only the Al states. However,
since H has no core electrons, we do use small H spheres. In this case,
it may be expected that the $s$ projection in the H sphere would represent
primarily the H $s$ contribution. Integrating the H $s$ projection
of the DOS over the occupied valence band, it is found that
there are 0.54 $e$ per H inside 1.1 $a_0$
LAPW spheres and 0.71 $e$ per H inside 1.3 $a_0$ spheres. So the
total DOS, the 1.3 $a_0$ H projection and the 1.1 $a_0$ projection
are in a ratio of 12 : 4.3 : 3.2, which is reasonably close to
12 : 3.6 : 2.7 for six isolated Watson sphere stabilized H$^-$ ions,
and lends support to the present ionic view of the bonding.
This is similar to the arguement that was used to support an ionic picture
of the bonding in NaAlH$_4$. \cite{aguayo}

In other materials, core level shifts can be used to evaluate
the ionic character. This is not generally possible in hydrides
because H has no core levels. However, as a further test, we
performed a calculation with the position of
one Li in the unit cell (6 Li and 12 H total)
exchanged with a H. In either a covalent or an ionic picture this
would be highly energetically unfavorable, both because of the 
unfavorable bond length, and because of the breaking of bonds
in the covalent case, and the placing of Li on an anion site in the
ionic case. Indeed it is found that this is highly unfavorable (by 0.76 Ry,
unrelaxed). However, significantly, the 1$s$ core level of the substituted
Li is at 4.1 eV higher binding energy (lower energy) than the other five
Li ions in the cell. This large shift,
which reflects the Coulomb potential,
clearly shows that the H site is an anion site.

The reason for the ionic electronic structure of Li$_3$AlH$_6$
can be understood as due to the long range Coulomb interaction
in solids. This Ewald contribution to the energy favors
ionic electronic structures, and is well known to stabilize
O$^{2-}$ in metal oxides, for example, even though dimers and
small molecules with the same metal - O neighbors may be
covalent. Here H$^{-}$ is stabilized in this way.

I am grateful for helpful conversations with
A. Aguayo, M. Gupta, R. Gupta and K. Yvon.
Work at the Naval Research Laboratory is
supported by the Office of the Naval Research.
The DoD-AE code was used for some calculations.

\begin{table}[tbp]
\caption{Experimental atomic coordinates for Li$_3$AlH$_6$ (Brinks and Hauback)
compared with the results of structural
relaxation within the LDA keeping the lattice vectors fixed at the
experimental values. These coordinates are given for the rhombohedral
setting, lattice vectors (5.6356,-0.07162,-0.07162),
(-0.07162,5.6356,-0.07162) and (-0.07162,-0.07162,5.6356) in \AA.}
\begin{tabular}{lddd}
Atom & $x$ & $y$ & $z$ \\
\hline
Li (EXP)  & 0.5595 & 0.0651 & 0.2487 \\
Al1 (EXP) & 0.0000 & 0.0000 & 0.0000 \\
Al2 (EXP) & 0.5000 & 0.5000 & 0.5000 \\
H1 (EXP)  & 0.0731 & 0.2950 & 0.9340 \\
H2 (EXP)  & 0.4138 & 0.2080 & 0.5482 \\
\hline
Li (LDA)   & 0.5627 & 0.0695 & 0.2486 \\
Al1 (LDA)  & 0.0000 & 0.0000 & 0.0000 \\
Al2 (LDA)  & 0.5000 & 0.5000 & 0.5000 \\
H1 (LDA)   & 0.0707 & 0.2960 & 0.9323 \\
H2 (LDA)   & 0.4118 & 0.2048 & 0.5496
\end{tabular}
\label{table1}
\end{table}

\end{multicols}
\end{document}